\documentclass[preprint,showpacs,preprintnumbers,amsmath,amssymb]{revtex4}

\usepackage{graphicx}
\usepackage{dcolumn}
\usepackage{bm}
\renewcommand{\vec}[1]{\boldsymbol{#1}}
\newcommand{\lb}{\mathrm{[}}
\newcommand{\rb}{\mathrm{]}}
\newcommand{\h}[1]{\hat{#1}}

\begin{document}

\preprint{HIP-2009-02/TH}

\title{Dirac Quantization Condition for Monopole\\ in Noncommutative Space-Time}

\author{Masud Chaichian$^{1,2}$}\email{masud.chaichian@helsinki.fi}
\author{ Subir Ghosh$^3$}\email{subir_ghosh2@rediffmail.com}
\author{Miklos L\aa ngvik$^1$}\email{miklos.langvik@helsinki.fi}
\author{Anca Tureanu$^{1,2}$}\email{anca.tureanu@helsinki.fi}
\affiliation{$^1$Department of Physics, University of Helsinki, P.O.
Box 64, FIN-00014 Helsinki, Finland}
 \affiliation{$^2$Helsinki
Institute of Physics, P.O. Box 64, FIN-00014 Helsinki, Finland}
\affiliation{$^3$Physics and Applied Mathematics Unit, Indian
Statistical Institute,  Kolkata-700108, India}


\begin{abstract}
Since the structure of space-time at very short distances is
believed to get modified possibly due to noncommutativity effects
and as the Dirac Quantization Condition (DQC), $\mu e =
\frac{N}{2}\hbar c$,  probes the magnetic field point singularity, a
natural question arises whether the same condition will still
survive. We show that the DQC on a noncommutative space in a model
of dynamical noncommutative quantum mechanics remains the same as in
the commutative case to first order in the noncommutativity
parameter $\theta$, leading to the conjecture that the condition
will not alter in higher orders.


\end{abstract}

\pacs{14.80.Hv, 02.40.Gh, 03.65.-w}
\maketitle

\section{Introduction}

The idea of magnetic monopoles - the so-far hypothetical particles
carrying magnetic charge - is one of the most influential in modern
theoretical physics. The first effective theoretical proposal that
magnetic charge should exist was made by Dirac \cite{Di}, who argued
that in quantum mechanics the unobservability of phase permits
singularities which manifest themselves as sources of magnetic
fields. The Dirac Quantization Condition (DQC), $\mu e =
\frac{N}{2}\hbar c$, is a topological property, independent of
space-time points, that tells us that the mere existence of a single
magnetic monopole would imply that the electric charge is quantized.
Before Dirac, the surprising asymmetry in Maxwell's equations made
Poincar\'e and J. J. Thompson introduce the magnetic charge in the
theory as an artefact for simplifying the computation, while P.
Curie suggested even the actual existence of magnetic charge
\cite{curie}. The idea of magnetic monopoles was later extended by
the discovery of monopole solutions of classical non-Abelian
theories \cite{WuYa,HoPoY} and the introduction of the concept of
dyons - particles carrying both electric and magnetic charge
\cite{SchJZPrSo}, and with its significant influence, eventually
leading to the concept of duality \cite{MO} and on the string
theory.

In  recent years magnetic monopole structures have created a lot of
interest in condensed matter physics. In studying the Anomalous Hall
Effect, magnetic monopole structures in {\it momentum space} have
been experimentally observed and theoretically explained in
\cite{nag,fang}.

However, till date no magnetic monopole has been found (for a
comprehensive review of magnetic monopole searches, see e.g.
\cite{gia}), but the theoretical interest has stayed, as nothing in
the theory has ever been found to contradict the DQC.

In the recent decade there has been a growing interest in the
research concerning noncommutative spaces mainly due to the results
in  \cite{Dop} and \cite{SeWi}. In  \cite{SeWi} it was shown that
string theory in a constant background field leads to a
noncommutative field theory as a low energy limit.

Moreover, the result of \cite{Dop} has encouraged many to believe
that noncommutative field theory is a step towards a more complete
description of physics. In the ''gedanken experiment'' of
\cite{Dop}, it was argued that in the process of measurement of
space points, as the energy grows, eventually black holes are formed
and consequently objects of smaller extent than the diameter of the
black holes cannot be observed and one can think of space-time
''points'' as operators obeying a Heisenberg-like uncertainty
principle from which it follows that space-time is homogeneous and
can be interpreted as being noncommutative.

Although the main interest in this field lies in the formulation of
a consistent field theory on a noncommutative space-time, it is also
interesting to apply the noncommutative space to pure quantum
mechanics to see whether it is possible to extend ordinary quantum
mechanics to the noncommutative case. Specifically, the result
\cite{Dop} is quantum mechanical in nature and some results such as
the DQC \cite{Di}, which we will be exploring in this Letter, are
not obtained directly from field theory.

In the noncommutative case, the space-time is particularly
sensitive to the short-distance effects. Since the DQC in its
essence probes the singularity structure of the magnetic
field, one would think that this condition could no longer
remain valid in the noncommutative case. This is the main
motivation for the present work.

We shall start by briefly reviewing one known method for deriving
the DQC in the commutative case. Then, starting from a classical
Lagrangian corresponding to a dynamical model of noncommutative
quantum mechanics, we shall derive the DQC to first order in the
noncommutativity parameter $\theta$, and finally we shall discuss
the result and its possible generalizations.

\section{ One way of deriving the DQC}

In the commutative case, there is an ingenious way to derive the
DQC, first introduced by Jackiw \cite{Jack},  which uses a
gauge-invariant algebra, dependent only on the magnetic field. The
derivation is an example of the three-cocycles which appear when a
representation of a transformation group is nonassociative; in
particular, when the translations group is represented by
gauge-invariant operators in the presence of a magnetic monopole,
the Jacobi identity among the translation generators fails. The
restoration of the associativity of {\it finite} translations leads
to the DQC. A sketch of the derivation \cite{Jack} will be quite
illuminating: for a nonrelativistic particle with the electric
charge $e$, moving in a magnetic field $\vec B(\vec{x})$, one starts
by finding the non-canonical quantum brackets
\begin{eqnarray}
\lb x_i, x_j \rb & = & 0, \hspace{15pt} \lb x_i, \pi_j \rb = i\hbar\delta_{ij}, \\
\lb \pi_i, \pi_j \rb & = &
i\hbar\frac{e}{c}\epsilon_{ijk}B_k(\vec{x}) \label{Jalg},
\end{eqnarray}
where one defines the operators $\pi_i$ in the $x$-representation as
\begin{equation}
\pi_i = -i\hbar\partial_i - \frac{e}{c}A_i(\vec{x}), \label{comrep}
\end{equation}
with $A_i(\vec{x})$ being the vector potential {\footnote{When deriving the DQC using gauge-variant vector
potentials, one still defines the magnetic field as $\vec{B} =
\nabla\times\vec{A}$ even though this cannot now remain true in the
whole space without the potential having singularities \cite{WuYa}.
Instead one may consider the vector potential as being defined in
many overlapping regions of space connected by singularity-free
gauge transformations \cite{WuYa}.}}.
These commutation relations, along with the Hamiltonian
\begin{equation}
H = \frac{\vec{\pi}^2}{2m}, \hspace{25pt} \vec{\pi} =
m\vec{\dot{x}},
\end{equation}
yield the well-known Lorentz-Heisenberg equations of motion
\begin{eqnarray}
\vec{\dot{x}} & = & \frac{i}{\hbar}\lb H, \vec{x} \rb = \frac{\vec{\pi}}{m}, \\
\vec{\dot{\pi}} & = & \frac{i}{\hbar}\lb H, \vec{\pi} \rb =
\frac{e}{2mc}\lb\vec{\pi}\times \vec{B} - \vec{B}\times\vec{\pi}\rb,
\end{eqnarray}
where $\vec \pi$ is the gauge-invariant mechanical momentum. So far
there is no restriction on $\vec {B}$ but the following  Jacobi
identity violation,
\begin{equation}
\frac{1}{2}\epsilon_{ijk}[[\pi_i, \pi_j], \pi_k] =
\frac{e\hbar^2}{c}\nabla\cdot \vec{B},
\end{equation}
indicates that  the magnetic field has to be   source-free.
Otherwise, $\nabla\cdot \vec{B} \neq 0$ will lead to a loss of
associativity of the translation operators $ T(\vec{a}) \equiv
\exp\Big(-\frac{i}{\hbar}\vec{a}\cdot\vec{\pi}\Big)$,
\begin{equation}
\Big(T(\vec{a_1})T(\vec{a_2})\Big)T(\vec{a_3})  =
\exp\Big(-\frac{ie}{\hbar c}\omega(\vec{x}; \vec{a}_1, \vec{a}_2,
\vec{a}_3)\Big)\times
T(\vec{a_1})\Big(T(\vec{a_2})T(\vec{a_3})\Big).
\end{equation}
Here $\vec{a}_i$ are  constant vectors and the non-trivial phase
factor turns out to be the magnetic flux coming out of the
tetrahedron formed by $\vec{a}_i$:
\begin{equation}
\exp\Big(-\frac{ie}{\hbar c}\omega(\vec{x}; \vec{a}_1, \vec{a}_2.
\vec{a}_3)\Big), \label{phaseJ}
\end{equation}
which is nonzero if a magnetic monopole is enclosed by the
tetrahedron. The phase factor \eqref{phaseJ} becomes 1 and thus the
associativity of {\it finite} translations in the presence of the
magnetic monopoles can be re-established for
\begin{equation}
\int d^3x\ \nabla\cdot \vec{B} = 2\pi\frac{\hbar c}{e}N,
\end{equation}
where $N$ is an integer. This condition, together with the Gauss
equation for a monopole of magnetic charge $\mu$, $ \nabla\cdot
\vec{B} = 4\pi \mu\delta^3(\vec{x}) $, yields the celebrated DQC
\begin{equation}\label{dqc}
\mu e = \frac{1}{2}N\hbar c.
\end{equation}
Note that  the Jacobi identity  is still violated at the location of
each monopole and these points are conventionally excluded from the
manifold.

\section {The noncommutative DQC}

The extension of the approach in \cite{Jack} to the noncommutative
case can be achieved once one finds the algebra of coordinate and
gauge-invariant momentum operators for a charged quantum mechanical
particle in motion in a magnetic field in the noncommutative
space-time, i.e. the analogue of the non-canonical algebra
\eqref{Jalg}. It is expected that the noncommutativity of space-time
coordinates would change the dynamics of the charged particle in the
magnetic field (i.e. the Lorentz force), and this in turn will
require a change in the commutation relations \eqref{Jalg}. However,
we can find the new noncommutative algebra by starting from a
classical Lagrangian, for example the one for the model
\cite{DuvHor}, and deriving the corresponding Dirac brackets and
then quantizing them. We therefore consider a Lagrangian of the form
\begin{equation}
L = \Big(P_i + \frac{e}{c} A_i\Big)\dot{X}_i -
\frac{1}{2}\epsilon_{ijk}P_i\dot{P}_j\theta_k -
\frac{1}{2m}\vec{P}^2 + eA_0, \label{lag}
\end{equation}
where $P_i$ is the momentum, $\theta_k$  - the noncommutativity
parameter, of dimension (length)$^2$/action, and $A_i, A_0$ - the
magnetic and electric potential, respectively. The Lagrangian
\eqref{lag} is a straightforward generalization to three-dimensions
of the one considered in \cite{LuStZa}, which is a Lagrangian for
the model \cite{DuvHor}.

The Lagrangian \eqref{lag} is a Lagrangian of dynamical
noncommutativity of the space coordinates. This claim is better
understood once we derive the Dirac brackets from this Lagrangian.
For this, we need the canonical momenta which are given by
\begin{equation*}
\pi_i = \frac{\partial L}{\partial \dot{X}_i} = P_i + \frac{e}{c}A_i,
\hspace{10pt} \pi_i^P = \frac{\partial L}{\partial \dot{P}_i} =
\frac{1}{2}\epsilon_{ijk}P_j\theta_k.
\end{equation*}
These lead to the constraints
\begin{equation*}
\eta_1 \equiv \pi_i - P_i - \frac{e}{c}A_i, \hspace{10pt} \psi_i
\equiv \pi^P_i - \frac{1}{2}\epsilon_{ijk}P_j\theta_k.
\end{equation*}
In the classical framework, with $\{X_i, \pi_j\} = \delta_{ij},
\hspace{1pt} \{P_i, \pi^P_j\} = \delta_{ij}$, we calculate the
constraint algebra,
\begin{eqnarray}
\{\eta_i, \eta_j\} & = & \frac{e}{c}(\partial_iA_j - \partial_jA_i) = \frac{e}{c}F_{ij} = \frac{e}{c}\epsilon_{ijk}B_k, \nonumber \\
\{\psi_i, \psi_j\} & = & -\epsilon_{ijk}\theta_k, \hspace{15pt} \{\eta_i, \psi_j\} = -\delta_{ij}. \label{abe}
\end{eqnarray}
From this algebra we find that the constraints are second class and,
performing the Dirac constraint analysis \cite{Di2, HenGit}, we
obtain the classical Dirac brackets as
\begin{eqnarray}
\{X_i, X_j\} & = & \frac{\epsilon_{ijk}\theta_k}{1 - \frac{e}{c}\vec{\theta}\cdot\vec{B}}\nonumber\\ \{X_i, P_j\}& =& \frac{\delta_{ij} - \frac{e}{c} B_i\theta_j}{1 - \frac{e}{c}\vec{\theta}\cdot\vec{B}}, \nonumber \\
\{P_i, P_j\} &=& \frac{\epsilon_{ijk}\frac{e}{c}B_k}{1 -
\frac{e}{c}\vec{\theta}\cdot\vec{B}}. \label{alg1}
\end{eqnarray}

This is exactly how interactions have been introduced in the model
of \cite{LuStZa}. In this model the noncommutativity of coordinate
operators is dynamical in the sense that it is generated within the
system. Thus the gauge field cannot be affected by the
noncommutativity which emerges upon quantization. Therefore the
field $A_i$ is the Abelian $U(1)$ gauge field in this model of
noncommutativity.

Our next step is to quantize the brackets. We do this by promoting
the classical variables $X_i$ and $P_i$ in the Dirac brackets
\eqref{alg1} to the status of operators $\h{X}_i$, $\h{P}_i$ and
multiplying the right-hand side of the Dirac brackets by $i\hbar$.
This is the standard procedure \cite{Di2, HenGit}. We consider the
Dirac brackets \eqref{alg1} expanded to first order in $\theta$ and
hereafter we perform all our calculations to this order only. We
resort to this approximation because we will need to find
representations for our operators in order to have a well-defined
quantum theory \cite{HenGit}, and this is a task that is difficult
to do exactly for the algebra \eqref{alg1}. The quantization of
\eqref{alg1} gives
\begin{eqnarray}
\lb\h{X}_i, \h{X}_j\rb & = & i\hbar\epsilon_{ijk}\theta_k + \mathcal{O}(\theta^2),  \label{alg2} \\
\lb\h{X}_i, \h{P}_j\rb & = & i\hbar\big[\delta_{ij} - \frac{e}{c}B_i(\h{\vec{X}})\theta_j
+ \frac{e}{c}\delta_{ij}\vec{\theta}\cdot\vec{B}(\h{\vec{X}})\big] + \mathcal{O}(\theta^2), \nonumber \\
\lb\h{P}_i, \h{P}_j\rb & = &
i\hbar\frac{e}{c}\epsilon_{ijk}B_k(\h{\vec{X}})\big[1 +
\frac{e}{c}\vec{\theta}\cdot\vec{B}(\h{\vec{X}})\big] +
\mathcal{O}(\theta^2). \nonumber
\end{eqnarray}
The algebra \eqref{alg2} poses a twofold problem. Firstly, the
operator $\h{P}_j$ in \eqref{alg2} does not represent the
translation generator, since there are extra terms on the right-hand
side of $\lb\h{X}_i, \h{P}_j\rb$, other than $i\hbar\delta_{ij}$.
Secondly, we face the problem of how to represent the operators
$\h{X}_i$, since they do not commute to first order in $\theta$.
This problem becomes much simpler if we are able to define some new
operators $x_i$, in terms of the old ones $\h{X}_i$ and $\h{P}_i$,
such that they commute to first order in $\theta$. An appropriate
definition for our purpose is
\begin{equation}
x_i = \h{X}_i + \frac{1}{2}\epsilon_{ijk}\h{P}_j\theta_k. \label{xdef} \\
\end{equation}
Then the functions of the operator $\h{X}_i$ can be expanded in
terms of the new coordinate operator $x_i$ as, e.g.,
\begin{equation}
B_i(\h{\vec{X}}) = B_i(\vec{x}) - \frac{1}{2}\epsilon_{njk}\theta_k\h{P}_j\partial_nB_i(\vec{x}) + \mathcal{O}(\theta^2). \label{funcdef}
\end{equation}
We use the operator $x_i$ \eqref{xdef} and the expansion
\eqref{funcdef} to obtain an intermediate algebra of $x_i$ and $\hat
P_j$, and further define the generator of translations corresponding
to $x_i$:
\begin{equation}
p_j = \h{P}_j -
\frac{1}{2}\frac{e}{c}\Big(\h{P}_j(\vec{B}\cdot\vec{\theta}) -
\h{\vec{P}}\cdot \vec{B}\,\theta_j\Big). \end{equation}
The newly-defined operators $p_i$ and $x_i$ obey the algebra
\begin{eqnarray}
\lb x_i, x_j\rb & = & 0 + \mathcal{O}(\theta^2)\nonumber\\\lb x_i,
p_j\rb &=& i\hbar\delta_{ij} + \mathcal{O}(\theta^2),
\label{alg3}\\
\lb p_i, p_j\rb & = & i\hbar\frac{e}{c}\epsilon_{ijk}B_k -
\frac{e}{2c}\Big[i\hbar\Big(p_{[j} \partial_{i]}
(\vec{B}\cdot\vec{\theta})+ p_{[i}\theta_{j]}\nabla\cdot {\bf B} +
{\bf p}\cdot\theta_{[i}\partial_{j]}{\bf B}\Big)\nonumber \\&& +
p_{[j}\lb p_{i]}, {\bf B} \rb\cdot{\vec\theta}  +\vec p\cdot\lb \vec
B,p_{[i}\rb\theta_{j]}\Big]  + \mathcal{O}(\theta^2), \nonumber
\end{eqnarray}
where the indices in brackets are anti-symmetrized.

To have properly quantized the algebra \eqref{alg3}, we need a
representation of its operators. From the similarity of the algebras
\eqref{alg3} and \eqref{Jalg}, we infer that in the
$x$-representation, we can realize the translation generators as
\eqref{comrep} plus an extra term involving the first order
noncommutativity contribution. Explicitly,
\begin{equation}
p_i = -i\hbar\partial_i - \frac{e}{c}A_i(\vec{x}) + T_i(\theta,\vec{x}) + \mathcal{O}(\theta^2).  \label{try}
\end{equation}
Inserting \eqref{try} into the commutator $\lb p_i, p_j\rb$ of
the algebra \eqref{alg3}, it simplifies to
\begin{equation}
\lb p_i, p_j \rb =  i\hbar\frac{e}{c}\epsilon_{ijk}B_k +
\frac{1}{2}\frac{e}{c}\Big[(i\hbar\partial_{[i} +
\frac{e}{c}A_{[i})\theta_{j]}\nabla\cdot\vec B\Big]+
\mathcal{O}(\theta^2). \label{ppbra}
\end{equation}
By directly computing the commutator of the operators $p_i$ in the
representation \eqref{try}, we have to reproduce the result
\eqref{ppbra}, which holds true if we set
\begin{eqnarray}
T_i(\theta,\vec{x}) & = & -\frac{1}{2}\frac{e}{c}\theta_i\nabla\cdot\vec B + G_i, \label{T} \\
\mathrm{where} \hspace{10pt} \partial_jG_i & = &
\frac{1}{2\hbar}\Big(\frac{e}{c}\Big)^2A_j\theta_i\nabla\cdot\vec B.
\label{bou}
\end{eqnarray}
Thus, the quantized algebra \eqref{alg3} is given by
\begin{eqnarray}
\lb x_i, x_j\rb & = & 0 + \mathcal{O}(\theta^2),\nonumber\\
\lb x_i, p_j\rb &=& i\hbar\delta_{ij} + \mathcal{O}(\theta^2), \label{alg4} \\
\lb p_i, p_j\rb & = & i\hbar\frac{e}{c}\epsilon_{ijk}B_k +
\frac{e}{2c}\Big[(i\hbar\partial_{[i} +
\frac{e}{c}A_{[i})\theta_{j]}\nabla\cdot\vec B\Big] +
\mathcal{O}(\theta^2) \nonumber
\end{eqnarray}
in the $x$-representation.

We can now calculate the Jacobi identities of the algebra \eqref{alg4}, and find that
the only non-vanishing one is:
\begin{eqnarray}
\frac{1}{2}\epsilon_{ijk}\lb\lb p_i, p_j\rb, p_k\rb & = &
-\hbar^2\frac{e}{c}\nabla\cdot \vec B +
\frac{i\hbar}{2}\Big(\frac{e}{c}\Big)^2\epsilon_{ijk}\partial_k(A_i\theta_j\nabla\cdot
\vec B) + \mathcal{O}(\theta^2).\label{j-id}
\end{eqnarray}
Since the nonvanishing terms in the right-hand side of \eqref{j-id}
are proportional to $\nabla\cdot\vec{B}$, for a divergenceless
magnetic field there are no Jacobi indentity violations. However, if
the magnetic field is produced by monopoles, $\nabla\cdot\vec{B} =
4\pi \mu\delta^3(\vec{x})$, the Jacobi identity \eqref{j-id} is
violated, meaning nonassociativity of the translation generators
$p_i$.

We would like to remark at this point that although the Lagrangian
\eqref{lag} contains no magnetic sources,
the algebra \eqref{alg4} is valid whether the magnetic field is source-free or
not. The reason is simply that the Lorentz-force describes the
movement of electrically charged particles in a magnetic field, but
does not set any requirement on how the magnetic field is produced.

Thus in the noncommutative space we end up with a Jacobi identity
violation consisting of the original commutative space term plus a
$\theta$-dependent total-derivative term. Let us recall that DQC
appears in the commutative case \cite{Jack} through a volume
integration (see  (\ref{phaseJ})) over the tetrahedron formed by the
three translation vectors $ \vec{a}_1, \vec{a}_2, \vec{a}_3$. Now,
the $\theta $-term in (\ref{j-id}), being a total derivative, should
contribute at the boundary of the tetrahedron. However, this
contribution will be necessarily zero, because the integrand
contains the $\delta$-function coming from $\nabla\cdot\vec{B}=4\pi
\mu\delta^3(\vec{x})$, which has support only at the origin, i.e. on
the monopole. Hence these two features conspire to cancel the effect
of the $\theta $-term.  DQC remains unchanged in the presence of
spatial noncommutativity, since the argument for restoring the
associativity of the noncommutative translation operators goes
through in the same manner as in the commutative case \cite{Jack},
but now with the translation operators
\begin{equation}
T_{NC}(\vec{a}) =
\exp\Big(-\frac{i}{\hbar}\vec{a}\cdot\vec{p}\Big),\label{transl_op}
\end{equation}
generated by $\vec{p}$ as the element of the algebra \eqref{alg4},
valid to first order in $\theta$ and with the $x$-representation
\eqref{try}.

\section{Summary and discussion}

We have explicitly shown that the DQC \eqref{dqc} remains unaltered
in noncommutative space to first order in $\theta$. Based on the
structure of the classical algebra \eqref{alg1} and the
representation of the quantum algebra \eqref{try}, and also
considering the fact that the form of any topological correction is
strongly constrained, we conjecture that the DQC will hold true in
all orders in the noncommutative space-time. We intend to elucidate
this issue in the future.

It would be interesting to obtain the same kind of indication of a
DQC to first order in $\theta$ using a noncommutative non-Abelian
vector potential \cite{SeWi,Hay}, especially since a gauge-covariant
noncommutative Aharonov-Bohm effect has been formulated in
\cite{us}. However, this formulation gives the required phase factor
with the help of non-Abelian noncommutative Wilson lines which are
notoriously tedious to work with even to first order in $\theta$,
due to the path ordering appearing in the Wilson line. Therefore,
obtaining a possible DQC in this approach stands as a challenge for
the future.

Our conclusion is that the  DQC  remains unchanged  in the
noncommutative case, to the first order in $\theta $ and expectably
to all orders. This is of significance, since a vast amount of work
has
 been devoted to studying various effects of noncommutative
space only to the lowest order in $\theta $. Finally, we would like
to mention that our work reinforces similar topological results in
the noncommutative case for other nonperturbative monopole-,
soliton- and dyon-solutions \cite{HaHaGrNeCiSch}.

\section*{Acknowledgments}
We are indebted to Claus Montonen and Shin Sasaki for illuminating
discussions. A.~T.~ acknowledges Projects Nos. 121720 and 127626 of
the Academy of Finland.

\end{document}